\documentclass[11pt,twoside]{book}
\usepackage{konkolyproc2}
\usepackage{longtable}
\usepackage{amsmath,amssymb}
\usepackage{graphicx}
\usepackage{lscape}
\usepackage{index}
\usepackage{natbib}
\usepackage{bigdelim}
\usepackage{multirow}
\makeindex

\begin{document}
\pagestyle{myheadings}
\setcounter{equation}{0}\setcounter{figure}{0}\setcounter{footnote}{0}\setcounter{section}{0}\setcounter{table}{0}\setcounter{page}{1}
\markboth{O. Hanyecz \& R. Szab\'o}{Pop. synth. of RRL}
\title{Population synthesis of RR Lyrae stars in the original $Kepler$ and $K2$ Fields of View}
\author{Ott\'o Hanyecz$^{1,2}$ \& R\'obert Szab\'o$^1$}
\affil{$^1$MTA CSFK, Konkoly Observatory, Budapest, Hungary \\ $^2$E\"{o}tv\"{o}s Lor\'{a}nd University, Budapest, Hungary}

\begin{abstract}
It is interesting to ask what fraction of the total available RR Lyrae (RRL) sample that falls in the $Kepler$ and $K2$ Fields of View (FoV) is known or discovered. In order assess the completeness of our sample we compared the known RRL sample in the $Kepler$ and $K2$ fields with synthetic Galactic models. The Catalina Sky Survey RRL sample was used to calibrate our method. We found that a large number of faint RRL stars is missing from $Kepler$ and $K2$ fields. 
\end{abstract}

\section{Simulation of the original $Kepler$ and $K2$ FoVs}
The original $Kepler$ Mission has been highly succesful in providing unprecedented details of the light variation of RR Lyrae stars, as well as in discovering new dynamical phenomena in their pulsation. Almost each object shows unique and novel characteristics, therefore the total number of RR Lyrae stars in the $Kepler$ and $K2$ fields is of particular interest.

To estimate the expected number of RRL stars we used the {\sc TRILEGAL} \citep{girardi2005} and {\sc Besan\c{c}on} models \citep{robin2003}. To get a sense of the uncertainties of the simulated number of RRLs in the instability strip we shifted the original $Kepler$ and $K2$ FoVs by $\pm 0.5$ degrees in R.A. and DEC directions. We used these standard deviations to check the changes of the simulated numbers of RRLs in slightly different fields.

In order to understand the completeness of the observed $Kepler$ and $K2$ RRL samples, we moved the original $Kepler$ field along the same galactic lattitude where the Catalina Sky Survey (CSS) made observations down to $19.5^{m}$ \citep{drake2013}.

Since CSS covered some of the $K2$ fields (and discovered a large number of RRL stars), we could use these measurements to compare the simulations with much deeper observational results in the $K2$ fields.

\section{Results}
\begin{figure}[!ht]
\includegraphics[width=1.0\textwidth]{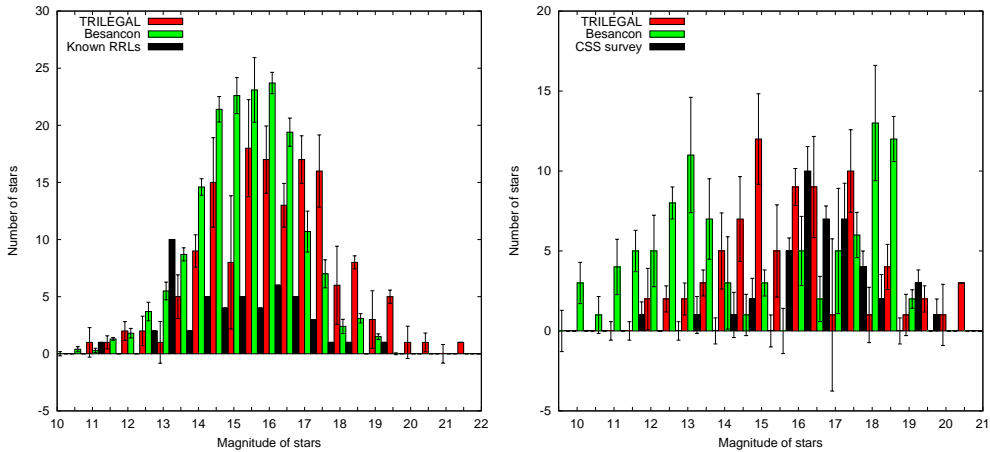}
\caption{The apparent magnitude distribution of the simulated stars and the known RRL stars in the original $Kepler$ FoV (left panel) and in $K2$ $Campaign$ $5$ (right panel). The red and green boxes show simulated stars from {\sc TRILEGAL} and {\sc Besan\c{c}on}, respectively. The black boxes show the 51 and 44 known RRLs in the $Kepler$, and in $K2$ $Campaign$ $5$, respectively.}
\label{fig:fig1}
\end{figure}

In Fig.~\ref{fig:fig1}. we plot the apparent magnitude distribution of the previously known RRLs and the simulated RRLs from {\sc TRILEGAL} and {\sc Besan\c{c}on} simulations for the original $Kepler$ FoV and $K2$ Campaign 5, respectively.

(1) We found good agreement between the simulated and observed RRL star numbers in a control field covered by CSS, which makes us confident that this method robustly estimates the number of RRL stars.

(2) We found 150 RRL in the $Kepler$ field in total and approximately the same amount in the $K2$ fields from {\sc TRILEGAL} simulations.

(3) We found good agreement between the {\sc TRILEGAL} sample and the observed RRL stars by CSS between $16^{m}$ and $20^{m}$ in the $K2$ fields.

(4) By accepting this calibration and using the predicted {\sc TRILEGAL} and {\sc Besan\c{c}on} numbers as a proxy, we predict that approximately three times more RRL stars should lurk in the $Kepler$ and $K2$ fields, than presently known.

These results will help us to better understand the RRL samples of the $Kepler$ fields, and motivates us to find the missing faint halo RRL population that could be used for galactic structure studies.

\section*{Acknowledgement}
This project has been supported by the NKFIH K-115709, the Lend\"ulet-2009 and LP2014-17 Young Researchers' Programs of the Hungarian Academy of Sciences, the ESA PECS Contract No. 4000110889/14/NL/NDe, the European Community's Seventh Framework Programme (FP7/2007-2013) under grant agreements no. 269194 (IRSES/ASK) and no. 312844 (SPACEINN).

\end{document}